\documentclass[]{article}
\usepackage[utf8]{inputenc}
\usepackage{graphicx}
\usepackage{caption}
\usepackage{subcaption}
\usepackage{siunitx}
\usepackage{braket}
\usepackage{float}
\usepackage{array}
\usepackage{multirow}
\usepackage{mathtools}
\usepackage{xcolor}
\usepackage{multicol}
\usepackage{geometry}
\usepackage{authblk}
\usepackage{comment}
\usepackage[numbers]{natbib}
\usepackage[colorlinks=true,
   linkcolor=blue,
   citecolor=blue,
   urlcolor =blue]{hyperref}

\newcommand{\SOne}{$^2$S$_{1/2}$ }
\newcommand{\DFive}{$^2$D$_{5/2}$}

\title{\bf{Coherence properties of highly-excited  motional states of a trapped ion}}

\author[]{V. Jarlaud\footnote{Present address: LP2N, Laboratoire Photonique, Numérique et Nanosciences, Université Bordeaux - IOGS - CNRS: UMR 5298, 1 rue François Mitterrand, 33400 Talence, France}}
\author[]{P. Hrmo\footnote{Present Address: Institute of Experimental Physics, University of Innsbruck, Innsbruck, Austria}}
\author[]{M. K. Joshi\footnote{Present Address: Institute for Quantum Optics and Quantum Information, Technikerstra{\ss}e 21a, 6020 Innsbruck, Austria}}
\author[]{R. C. Thompson}
\affil[]{Quantum Optics and Laser Science, Blackett Laboratory, Imperial College London, \\ Prince Consort Road, London SW7 2AZ, United Kingdom}
\date{}

\begin{document}

\maketitle

\begin{abstract}
We present a study of the  coherence properties of a variety of motional states of a single ion confined in a Penning ion trap.  We demonstrate that the motion of the ion has a coherence time of the order of one second, using Ramsey interferometry.  We introduce a technique for preparing the ion in an incoherent superposition of highly-excited motional states using a simple modification of optical sideband cooling.  Coherent manipulation of these states allow measurements of  optical and motional coherence to be carried out. We show that these highly-excited motional state superpositions have long coherence times despite the incoherent preparation of the states. Such states can be useful for sensitive motional dephasing measurements.

\end{abstract}

\section{Introduction}\label{intro}         
Isolated atomic systems have been shown to have remarkable properties enabling high fidelity quantum control, sensing, and precision measurements \cite{leibfried03, Degen2017, Meyer2001, Ludlow2015, Pezze2018}. Laser cooled trapped ions in an ultra-high vacuum environment are particularly well isolated from environmental perturbations and constitute an ideal platform to study quantum phenomena that require long observation times \cite{Kotler2014} and high-fidelity quantum control \cite{Blatt2012,Wineland2003}. A single trapped ion can also act as an almost ideal model of a quantum harmonic oscillator allowing for studies of fundamental quantum optics phenomena such as the generation of Schr{\"{o}}dinger cat states \cite{McDonnell2007,Monroe1996}, large coherent states \cite{Alonso2016}, phonon arithmetic \cite{Um2016} and NOON states \cite{Zhang2018}. Electromagnetically induced transparency (EIT) cooling and resolved sideband cooling methods \cite{Morigi2000a, Lechner2016, Diedrich1989} have made possible fast and efficient preparation of the ions' motional ground state. In turn, this allows the motional states of the ions to be manipulated using laser or microwave fields to perform high-fidelity quantum gates, which are the building blocks of trapped-ion-based quantum computation and simulation \cite{Blatt2012,Sorensen1999,Haffner2008,Bohnet2016}.

In the aforementioned applications, one of the effects limiting fidelity and accuracy of the desired quantum operations is the coherence of the electronic states. For a qubit stored in the internal states of the atomic species, the maximum coherence time is given by the excited state lifetime, while instability of the control laser/microwave fields or the external magnetic fields present in the laboratory can lead to dephasing, which further reduces it. To mitigate these effects for a qubit addressed on an optical transition, sophisticated techniques to achieve an ultra-stable laser frequency lock \cite{Kessler2012} and a high degree of isolation from external magnetic fields \cite{Bruzewicz2019} have been developed along with a host of dynamical decoupling schemes \cite{Wang2017,Biercuk2009}. A second effect that is particularly important for the fidelity of operations mediated by the motional states of the ion is the coherence of the phonon state \cite{turchette00}. In the ion trapping community this was typically investigated under the guise of ion `heating' \cite{Turchette2000}, which leads to a change in the phonon state by absorption of a phonon at the trap frequency. Extensive literature exists studying the origins of this phenomenon \cite{turchette00,brownnutt15,Boldin2017}, with ongoing efforts on mitigation strategies involving surface science \cite{hite13,Deslauriers2006,Chiaverini2014}, in-situ cleaning \cite{Allcock2011,McConnell2015} and cooling of the trap to cryogenic temperatures \cite{Labaziewicz2008,bruzewicz15,Lakhmanskiy2019}. However, analogous to the dephasing of the electronic state, the phonon state also undergoes dephasing when exposed to a noise source that alters the trap frequency. Measurements of the motional dephasing rate have often gone unreported in the literature as the overall motional coherence tended to be limited by the heating rate \cite{Schmidt-Kaler2003}. As the heating rates have improved over time due to the above mentioned studies, motional dephasing has become an important consideration for future experiments, in particular ones based on surface trap designs where there are many DC electrode voltages or radio-frequency amplitudes used to generate the trapping potential that have to be stabilised to prevent trap frequency fluctuations. 
\par
In this article, we investigate motional coherence in a ground-state cooled ion. We also extend the work by studying the motional coherence of high lying phonon states, which are populated by performing sideband heating of an ion, the opposite of the conventional sideband cooling method \cite{joshi19}. Using this technique, measurements of the motional dephasing can be performed with greatly reduced sensitivity to motional heating, providing an alternative to the coherent state scheme of Talukdar \textit{et al.} \cite{talukdar16}. Besides quantum information processing, other experiments that can benefit from detailed analysis of motional dephasing include force sensing \cite{Biercuk2010,mccormick19} and quantum logic spectroscopy \cite{Schmidt2005}.

The article is structured as follows. We  first give a brief description of the experimental set-up in section \ref{setup} and then present the results of experiments conducted on an ion prepared in the motional ground state (section \ref{GS}). This serves to introduce the techniques we use for coherently manipulating the motional state of the ion. We demonstrate, using a motional Ramsey sequence, that the motional coherence time near the ground state approaches one second. Then in section \ref{highN}, we show how we can prepare an ion in a high, non thermal, motional state using a technique similar to the well-known sideband cooling technique.  We demonstrate the capability to carry out coherent manipulations of the ion, despite the fact that it is in an incoherent mixture of motional states.  Ramsey interferometry of these states demonstrates motional coherence times similar to those found when starting from the ground state. The Ramsey interference pattern shows features due to the loss of optical and motional coherence on widely different timescales. Finally, in section \ref{conclusion} we present our conclusions.

\section{Experimental set-up}\label{setup}
The experiments presented in this article are carried out with a single $^{40}$Ca$^+$ ion confined in a Penning trap having a magnetic field of 1.89 T and a static quadrupole electric potential. The set-up has been described in previous articles \cite{mavadia14,hrmo19} and here we will only recall some of its features, important for our discussion. Using a combination of Doppler and resolved sideband cooling, we can prepare an ion in its motional ground state, both for the axial and the radial motions, respectively along and perpendicular to the magnetic field axis \cite{hrmo19,goodwin16}. In the following, we will however only consider the axial motion, which is harmonic, with a typical frequency in the range 100--450 kHz. The eigenstates of this quantum harmonic oscillator are denoted as $\ket{n}$.

\begin{figure}
    \centering
    \includegraphics[scale=0.5]{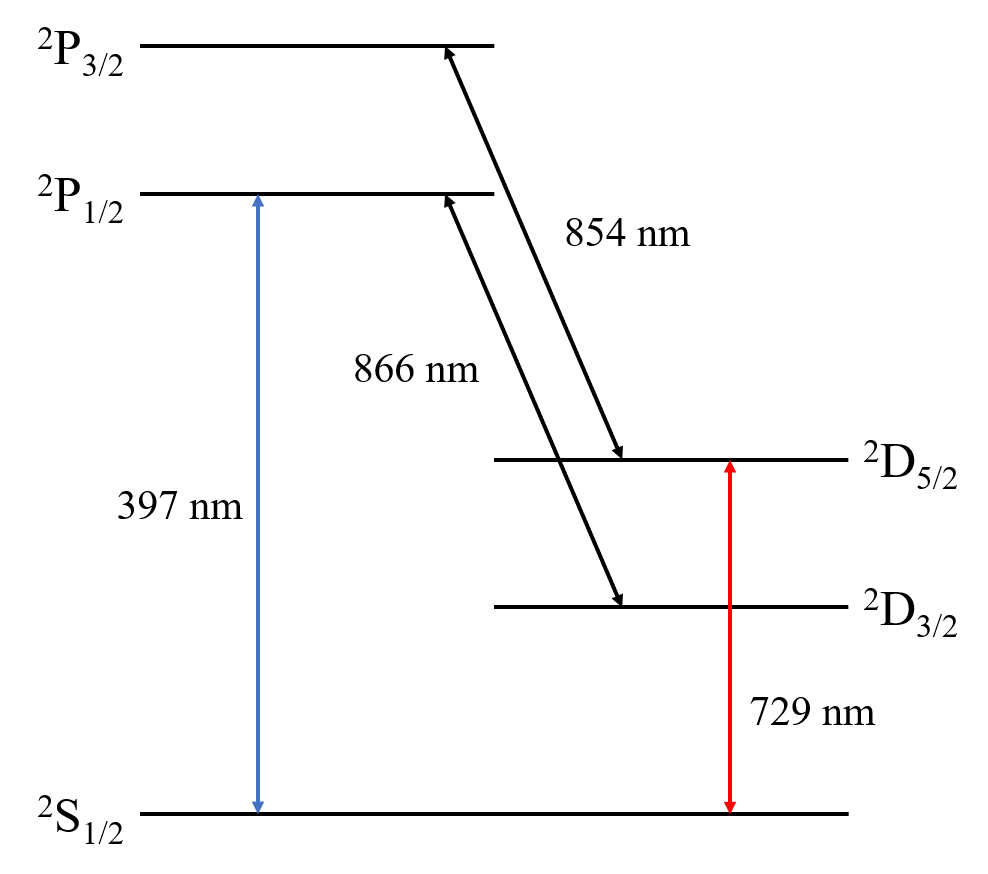}
    \caption{Partial energy structure of $^{40}$Ca$^+$ showing the wavelength of the transitions driven by lasers. The magnetic sub-levels are not represented.}
    \label{CaLevels}
\end{figure}

Doppler cooling is performed by driving the $^2$S$_{1/2}\leftrightarrow$ $^2$P$_{1/2}$ transition of $^{40}$Ca$^+$ at 397 nm (see figure \ref{CaLevels}) with a pair of red-detuned lasers while two ``repumper'' lasers at 866 nm and 854 nm empty the metastable $^2$D$_{3/2}$ and $^2$D$_{5/2}$ levels respectively, which may be populated by spontaneous decay from the $^2$P$_{1/2}$ level \footnote{The magnetic sub-levels of $^2$S$_{1/2}$ are separated by about $\SI{53}{\giga\hertz}$ in a magnetic field of 1.89 T. Therefore two different lasers at 397 nm are necessary to excite the $^2$S$_{1/2}\leftrightarrow$ $^2$P$_{1/2}$ transition. The repumper lasers are modulated by an electro-optic modulator to generate the various frequencies required to address the different magnetic sub-levels.}. De-excitation to the $^2$D$_{5/2}$ level is made possible by the strong magnetic field of the Penning trap ($j$-mixing) hence the necessity of using a laser at 854 nm \cite{crick10}. For sideband cooling, one Zeeman component of the narrow linewidth transition $^2$S$_{1/2}\leftrightarrow$ $^2$D$_{5/2}$ at 729 nm is addressed. The ``repumper'' laser at 854 nm is kept on during sideband cooling to excite the ion from $^2$D$_{5/2}$ to $^2$P$_{3/2}$ from where it quickly decays to $^2$S$_{1/2}$. This greatly increases the cooling rate and allows us to reach the motional ground state $\ket{n}=\ket{0}$ with a high probability ($>98 \%$) after a few milliseconds of cooling  \cite{goodwin16}.

The same $^2$S$_{1/2}\leftrightarrow$ $^2$D$_{5/2}$ laser at 729 nm is used to coherently drive the ion. In the context of coherent manipulations, we will consider the calcium ion as a two-level system with the electronic ground state \SOne in the sublevel $m_j=-1/2$ written as $\ket{g}$ and an excited state $\ket{e}$ corresponding to the metastable state \DFive, in the sublevel $m_j=-3/2$. The ion can be prepared in \SOne, $m_j=-1/2$ by optical pumping, by turning off one of the 397 nm lasers used for Doppler cooling. The decay rate from the metastable (excited) state to the ground state is $\Gamma=2\pi\times\SI{0.136}{\hertz}$ which is much slower than the optical decoherence rate, so spontaneous emission from this level can be neglected. The laser used to drive the transition between $\ket{g}$ and $\ket{e}$ is an external cavity diode laser which is locked to a high-finesse optical cavity.  The resulting linewidth is less than 1 kHz. The frequency of the laser can be fine-tuned thanks to an acousto-optic modulator (AOM). The AOM is driven by a direct digital synthesiser\footnote{Analog Devices AD9910} (DDS) which can change the frequency, phase and amplitude of the radio-frequency signal very rapidly. The total duration needed to change the frequency (or another parameter) of the laser will be determined by the clock rate of our control system (25 MHz), the switching time of the DDS (4 ns), propagation delays and the rise time of the AOM. Overall, this delay is less than a microsecond and is constant. It is important to note that the DDS operates in a phase continuous manner which means that when the frequency of the DDS is switched, there is no discontinuity in the radio-frequency signal, that is, the signal with the new frequency does not begin at a specific point of the sinusoid.\footnote{This may seem to be an unimportant technical detail but actually has consequences for motional Ramsey interferometry, as we will see in the following sections.}

The electronic state of the ion is determined by measuring the fluorescence from the Doppler cooling lasers in the absence of the repumper laser at 854 nm. If the ion is in the metastable state $\ket{e}$, this fluorescence is absent. In order to determine the probability of excitation to a sufficient precision, experimental sequences are repeated many times (typically 100 or 200 times) to create one data point.

\section{Motional coherence near the mechanical ground state}\label{GS}        
A coherent superposition of motional states can be created by applying resonant laser pulses on different sidebands. Starting with an ion cooled to the ground state, we can create, for example, an equal superposition of $\ket{g,0}$ and $\ket{g,1}$ by performing a $\pi/2$ pulse on the carrier followed by a $\pi$ pulse on the first red sideband. Applying a mirrored pulse sequence after a wait time $T$, during which the laser is off, completes a Ramsey experiment for the ion's motion. Similarly, a superposition of $\ket{g,0}$ and $\ket{g,2}$ can be realized by replacing the first pulse on the carrier by a pulse on the first blue sideband.

We can show (derivation in  Appendix \ref{math}) that after a pulse sequence like the one described above, the probability for the ion to be in the electronic excited state is given by:
\begin{align}
    P_e=\frac{1}{2}\left(1-V(T)\cos\left(\Delta\omega T+\Phi_l \right)\right)
    \label{proba0n}
\end{align}
where $\Delta\omega$ is the frequency difference between the targeted transitions (e.g. carrier and first red sideband) and $\Phi_l$ is a constant phase term which depends on the parameters of the laser pulses. The visibility $V(T)$ is a term that takes into account different effects which reduce the amplitude of the oscillations in $P_e$. We will discuss it below. The presence of the term $\Delta\omega T$ in equation \ref{proba0n} is due to the DDS being phase continuous. With a phase coherent system, this term would not appear. From equation \ref{proba0n}, we see that oscillations of the excitation probability can be observed by changing the wait time $T$ or the phase $\Phi_l$ which can be done for example by scanning the phase of the laser on the last pulse.

Ideally, for the simple experiment described above, the term $V(T)$ is constant and equal to one, in which case equation \ref{proba0n} yields oscillations with a visibility equal to one. In practice, several effects tend to reduce the visibility such as imperfect laser pulses (non-zero detuning, pulses too long or too short), instabilities in the motional frequency or heating of the ion whereby its motional state is changed. The latter two effects can be thought of as a loss of coherence of the ion's motion. 

\begin{figure}       
    \centering
    \includegraphics[scale=0.6]{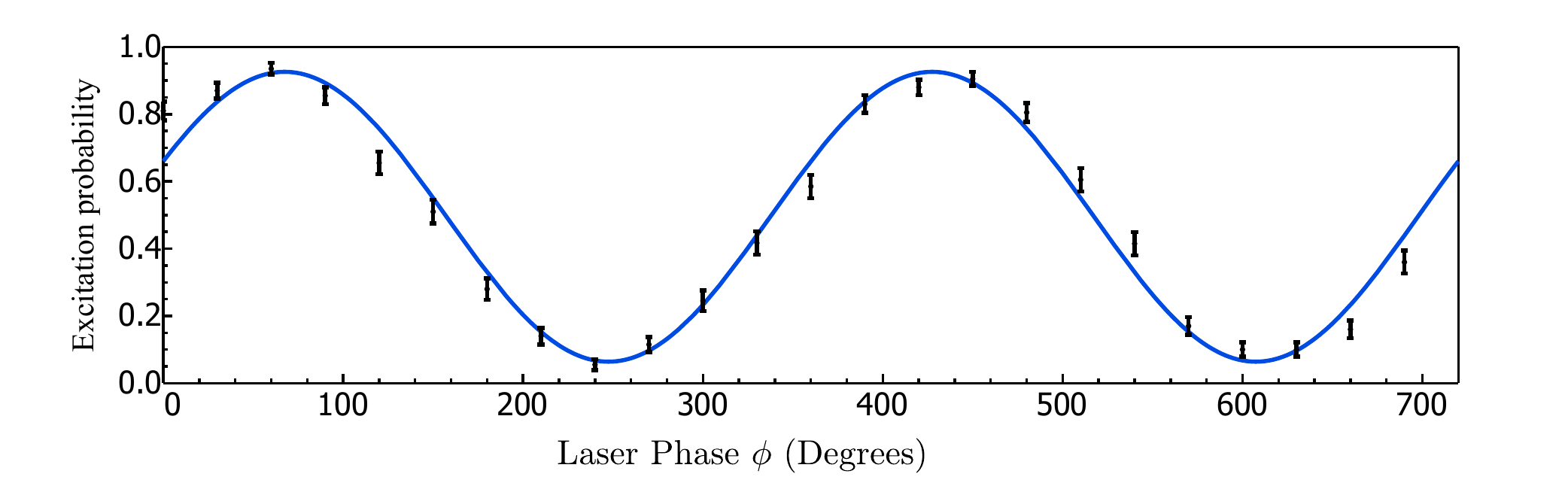}
    \caption{Probability of excitation as a function of the phase offset on the last pulse of the Ramsey sequence for a superposition of $\ket{g,0}$ and $\ket{g,1}$. The wait time in the sequence is 50 ms. Each data point is an average of 200 repeats.
    }
    \label{phaseScan}
\end{figure}

\begin{figure}      
    \begin{subfigure}{0.5\textwidth}
    \centering
    \includegraphics[scale=0.5]{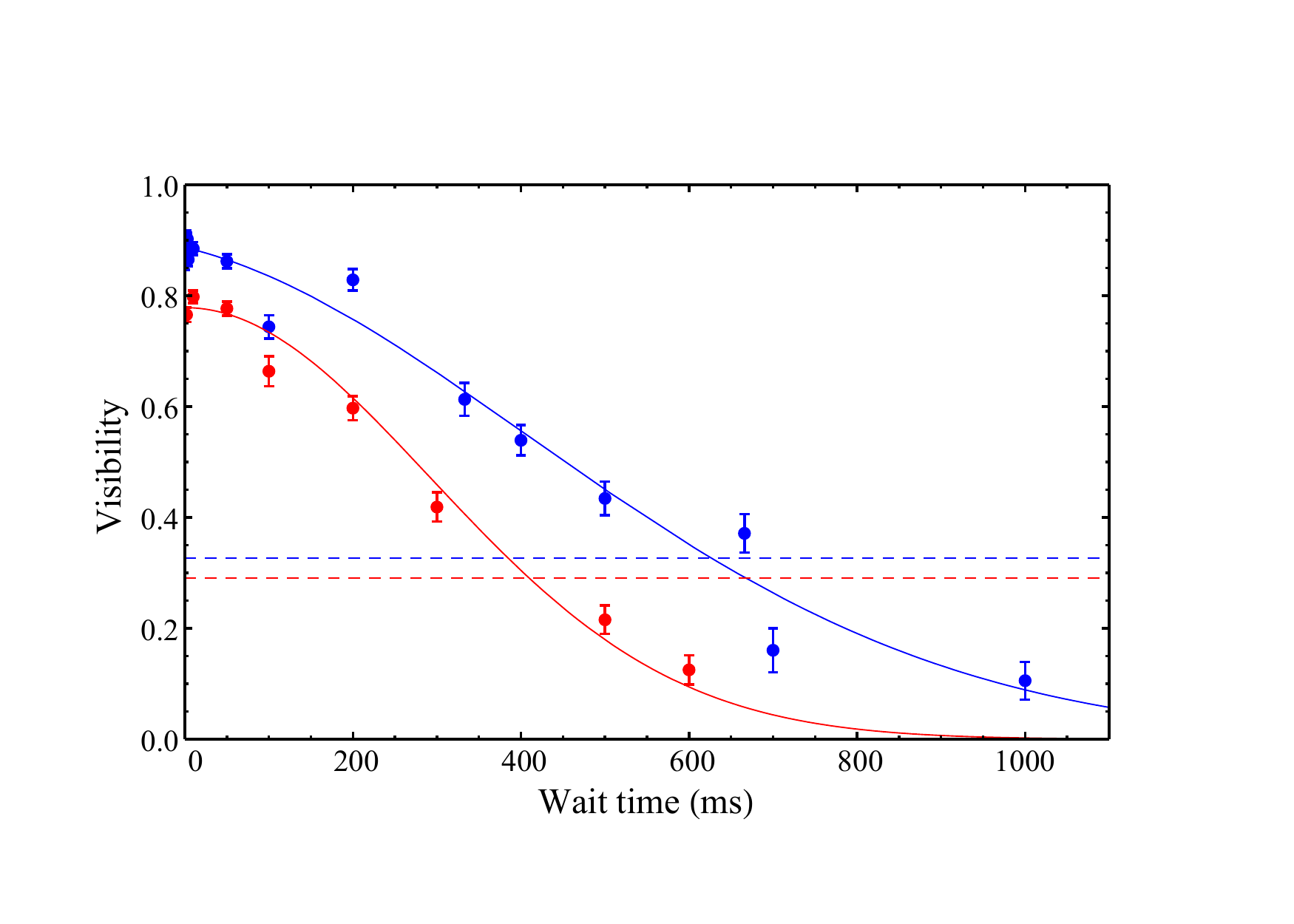}
    \caption{}
    \end{subfigure}
    \begin{subfigure}{0.5\textwidth}
    \includegraphics[scale=0.5]{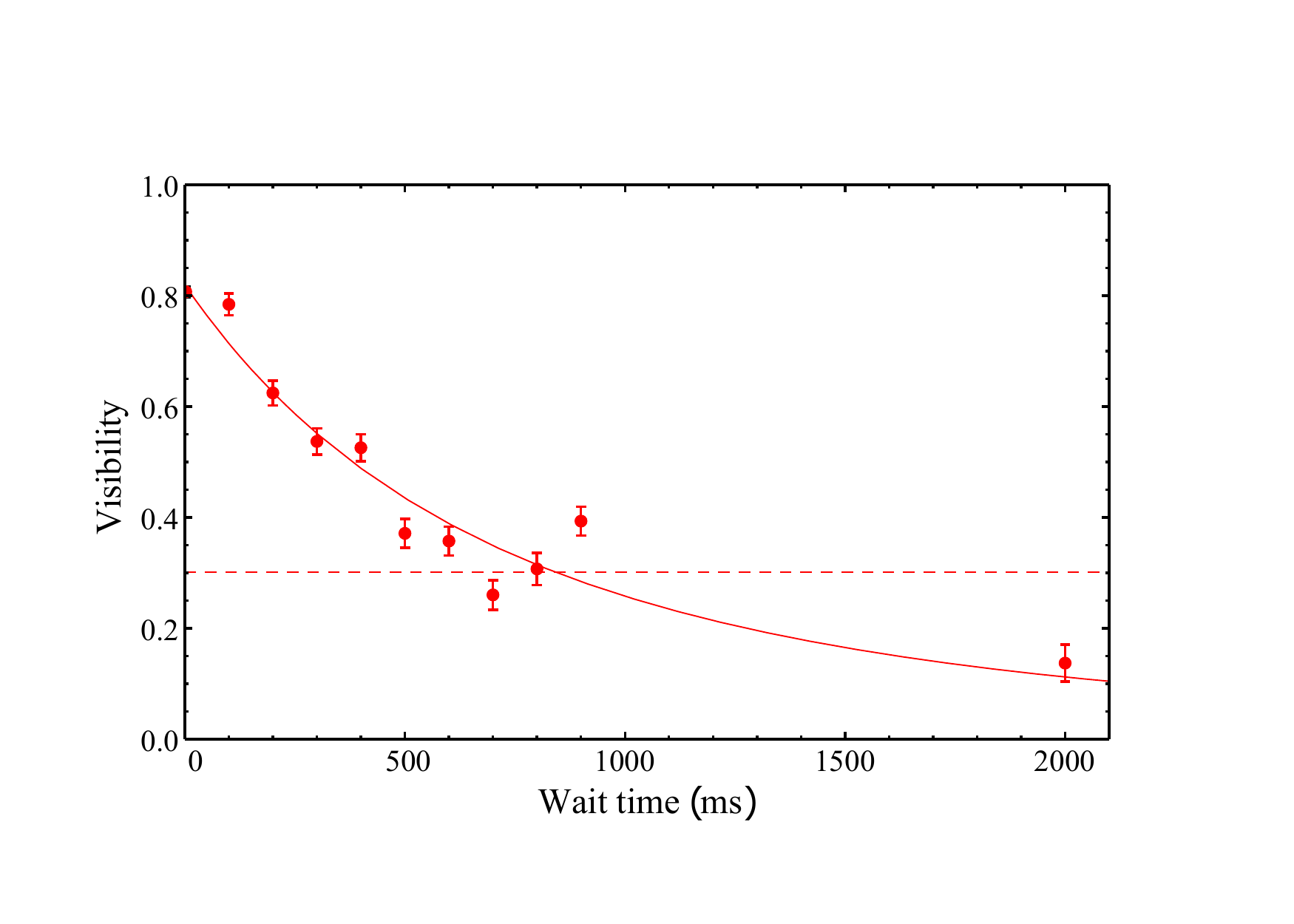}
    \caption{}
    \end{subfigure}
    \caption{\textbf{(a)} Visibility of phase scans (at $\nu_z=\SI{420}{\kilo\hertz}$) as a function of the wait time $T$ in the Ramsey sequence for a superposition of $\ket{g,0}$ and $\ket{g,1}$ (blue) and a superposition of $\ket{g,0}$ and $\ket{g,2}$ (red). The dashed lines mark the $V_0/e$ point where the coherence time is measured. The data points are fitted following equation \ref{visEq}, and the coherence times are around 0.6 s and 0.4 s for the blue and red curves respectively. \textbf{(b)} Visibility of phase scans as a function of the total wait time $T$ in a motion echo experiment on a superposition of $\ket{g,0}$ and $\ket{g,2}$. The coherence time is about 0.8 s.}
    \label{phaseScanVis}
\end{figure}

Figure \ref{phaseScan} gives an example of oscillations obtained from a phase scan. Here a wait time of 50 ms was inserted between the first two and last two laser pulses. The data points in figure \ref{phaseScanVis}(a) were obtained by performing Ramsey experiments with superpositions of $\ket{g,0}$ and $\ket{g,1}$ (blue curve) and $\ket{g,0}$ and $\ket{g,2}$ (red curve) for different wait times $T$. The figure shows the visibility of the oscillations decreasing as the wait time in the Ramsey sequence increases. Even with no wait time, the visibility is less than one. We can attribute this to imperfect laser pulses, whereas the loss of visibility with time is a consequence of decoherence.

The model we use for decoherence is based on earlier work \cite{turchette00} where coupling of a harmonic oscillator to amplitude and phase reservoirs was studied. The first effect to consider is heating of the ion motion which can be caused for instance by electrical noise at the ion's motional frequency. Heating, which changes the motional state of the ion (and therefore its energy), corresponds to an amplitude coupling and its effect on the motional Ramsey experiment is a reduction of the visibility of the oscillations.  This reduction depends on the heating rate $\Dot{\bar{n}}$ (which we assume constant); the time $T$ between the laser pulses; and the difference $\Delta n$ in motional state number between the two parts of the superposition, and it evolves as $\left(1+\Dot{\bar{n}}T\right)^{-(1+\Delta n)}$. The other effect leading to decoherence is the motional dephasing caused by fluctuations of the trap frequency due to instabilities of the trap's power supply.  Based on direct measurements of the supply voltage, we expect these fluctuations to be much slower than the length of a single experimental realisation. The effect is thus modeled as a static mismatch between the motional mode frequency and the frequency difference between the carrier and sideband pulses. Assuming this random frequency mismatch is normally distributed with a standard deviation $\sigma$, we obtain  the visibility as a function of the wait time between the pairs of laser pulses, which is given by:
\begin{align}
    V(T)=V_0\exp{\left(-\frac{1}{2}(\Delta n)^2\sigma^2T^2\right)}\left(1+\Dot{\bar{n}}T\right)^{-(1+\Delta n)}
    \label{visEq}
\end{align}
with $V_0$ the visibility for no wait time, which is not necessarily equal to one, as explained above. 
Using this equation to fit the data in figure \ref{phaseScanVis}, we find $\sigma=\SI{2.0(3)}{\radian\per\second}$, $\Dot{\bar{n}}=0.21(16)$ s$^{-1}$ and $V_0=0.89(1)$ for the superposition of $\ket{g,0}$ and $\ket{g,1}$ and $\sigma=\SI{1.5(2)}{\radian\per\second}$, $\Dot{\bar{n}}=\SI{0.22(17)}{\per\second}$ and $V_0=0.79(1)$ for the superposition of $\ket{g,0}$ and $\ket{g,2}$. Defining the coherence time $T_c$ as the point where the visibility is equal to $V_0/e$, we estimate $T_c(\Delta n=1)\approx\SI{0.6}{\second}$ and $T_c(\Delta n=2)\approx\SI{0.4}{\second}$. The values of the heating rate found with these experiments are in good agreement with what was measured via sideband thermometry in an earlier experiment: $\Dot{\bar{n}}=0.3(2)$ \cite{goodwin16}.

Because the heating rate in our trap is quite low, its contribution to motional decoherence can be hard to measure. In order to help distinguish it from dephasing, we perform the following experiment: in a manner similar to a spin echo for a superposition of electronic states, the dephasing in a superposition of motional states can be partially compensated by applying the equivalent of a ``$\pi$ pulse'' midway through the wait time of a Ramsey experiment. In this context, by $\pi$ pulse we mean swapping the populations in the two Fock states of the superposition, which is done with three laser pulses. By analogy with the spin echo, this technique is referred to as ``motion echo'' \cite{mccormick19bis}. For a superposition of $\ket{g,0}$ and $\ket{g,2}$, the population swapping is performed by applying a $\pi$ pulse on the first blue sideband flanked by two $\pi$ pulses on the first red sideband. A pulse sequence with one equivalent $\pi$ pulse removes the effect of a mismatch between the driving field difference frequency (i.e. between carrier and first red sideband) and the ion's motional frequency. It thus removes frequency drifts that are slower than the repetition rate of the experiment.
Higher order perturbations can however still contribute to dephasing.

Figure \ref{phaseScanVis}(b) shows the visibility of phase scans in motion echo sequences with different wait times on a superposition of $\ket{g,0}$ and $\ket{g,2}$. The data is fitted using the same model as for the Ramsey experiment above. We find $\sigma$ to be consistent with zero, $\Dot{\bar{n}}=0.47(6)$ and $V_0=0.82$. The coherence time is $T_c(\Delta n = 2)\approx\SI{0.84}{\second}$, a substantial increase compared to the Ramsey experiment without the equivalent $\pi$ pulse. The motion echo sequence effectively suppresses dephasing, leaving heating as the dominant source of decoherence. Although the heating rate found here is slightly higher than the values found with Ramsey experiments, it remains consistent with the results of sideband thermometry.

\section{Coherent manipulations in high motional states}\label{highN}
The experiments presented in section \ref{GS} relied on an initial preparation of the ion in a well defined Fock state. This was achieved via sideband cooling which allowed us to reliably prepare the ion in the motional state $\ket{0}$. From this point, we were able to create a motional superposition of $\ket{0}$ and $\ket{n}$. However, the method used in section \ref{GS} where we used two laser pulses only gives access to a limited range of superpositions. As the Rabi frequency of a sideband transition near the motional ground state rapidly decreases with the sideband order, we are effectively limited to using pulses on the carrier transition and the first order sidebands only.

One way of preparing an ion, initially in the ground state, in an arbitrary Fock state is to apply a series of many $\pi$ pulses, alternating between blue and red sidebands, in order to incrementally increase the motional state number. This approach was used, for instance, in ref. \cite{mccormick19} to create Fock states of up to $\ket{n=100}$. The limitation of this method is that it requires an excellent control of the laser pulses' parameters (duration, frequency, intensity) in order to have a full population transfer every time i.e. to perform perfect $\pi$ pulses. A small deviation from a perfect $\pi$ pulse quickly results in a diffusion of the motional state such that  a given Fock state $\ket{n}$ cannot be prepared with a good fidelity. This problem will naturally become more severe with an increasing number of laser pulses.

Here, we employ an alternative way of preparing an ion in a high motional state using a method analogous to sideband cooling where the ion is driven incoherently using laser light. We call this technique sideband heating in analogy  to sideband cooling but its effect is not a thermalisation of the ion. We make use of the variations of the motional sidebands' Rabi frequency with the motional state number $n$ in order to create a narrow distribution of states around particular points. This method does not necessarily yield a better preparation fidelity than a coherent drive approach but has the advantage of being very simple to implement with a much less stringent degree of control required while allowing  high motional states to be reached quickly. Coherent manipulations on the ion remain possible afterwards despite the incoherent nature of the sideband heating process.

\subsection{Sideband heating}
Sideband heating consists of increasing the motional state number $n$ (i.e. the energy) of the ion by driving a blue sideband with the $\SI{729}{\nano\meter}$ laser, while the $\SI{854}{\nano\meter}$ laser acts as a repumper. The connection with sideband cooling (described in \cite{goodwin16}) is apparent and it is performed experimentally in a similar manner. This technique is also described briefly in ref. \cite{joshi19} but we explain the general principle  below. We note that a similar scheme to create highly-excited Fock states was proposed in ref. \cite{Cheng2018}.

In order to understand the sideband heating technique, we need to consider the Rabi frequency $\Omega_{n,n+s}$ for the transition $\ket{g,n}\leftrightarrow\ket{e,n+s}$ which is given by:
\begin{align}
    \Omega_{n,n+s}=\Omega_0e^{-\eta^2/2}\eta^{\vert s\vert}\sqrt{\frac{n_<!}{n_>!}}L_{n_<}^{(\vert s\vert)}\big(\eta^2\big).
\label{rabiCoupEq}
\end{align}\\
where $\Omega_0$ is the carrier Rabi frequency for an ion at rest (not trapped); $n_>$ and $n_<$ are respectively the maximum and minimum of ${n}$ and ${n+s}$. $L_{n_<}^{\vert s\vert}(\eta^2)$ is a generalised Laguerre polynomial and $\eta$ is the Lamb-Dicke parameter defined by
\begin{align}
    \eta=\sqrt{\frac{\hbar k^2}{2m\omega_z}}
\end{align}
where $m$ is the mass of the ion and $k$ the wavevector of a photon at the frequency of the optical transition. The important consequence of equation \ref{rabiCoupEq} is that the Rabi frequency on a given sideband depends on the motional state of the ion and that for  certain values of the motional state, the Rabi frequency is almost zero. At these points, there is effectively no coupling between the ion and the laser light when tuned to the frequency of that sideband \cite{joshi19}.

\begin{figure}
    \centering
    \includegraphics[scale=1.2]{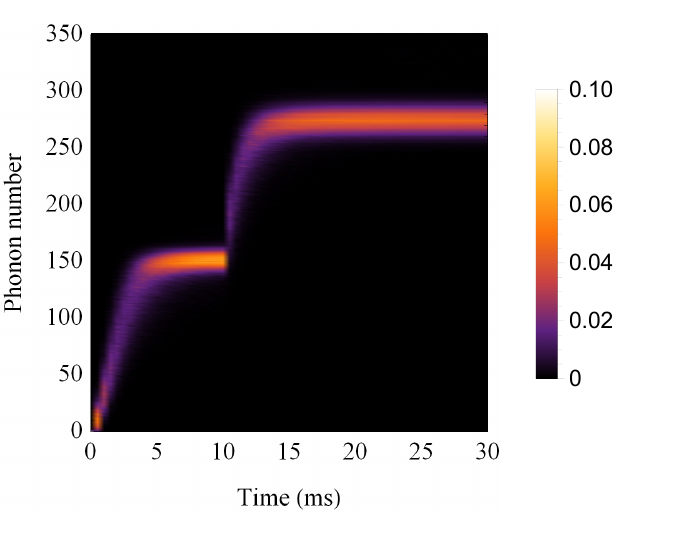}
    \caption{Simulated probability of motional state occupation as a function of time in a sideband heating sequence. During the first 10 ms, the first blue sideband is targeted, and the motional state of the ion concentrates around the zero-coupling point of this sideband at $n\approx 160$. During the last 20 ms, the second blue sideband is targeted, and the ion's motional state rises to higher phonon numbers until it reaches the zero-coupling point of this sideband at  $n\approx 290$. }
    \label{twoDPlot}
\end{figure}

\begin{figure}
\begin{subfigure}{0.5\textwidth}
\centering
\includegraphics[scale=0.55]{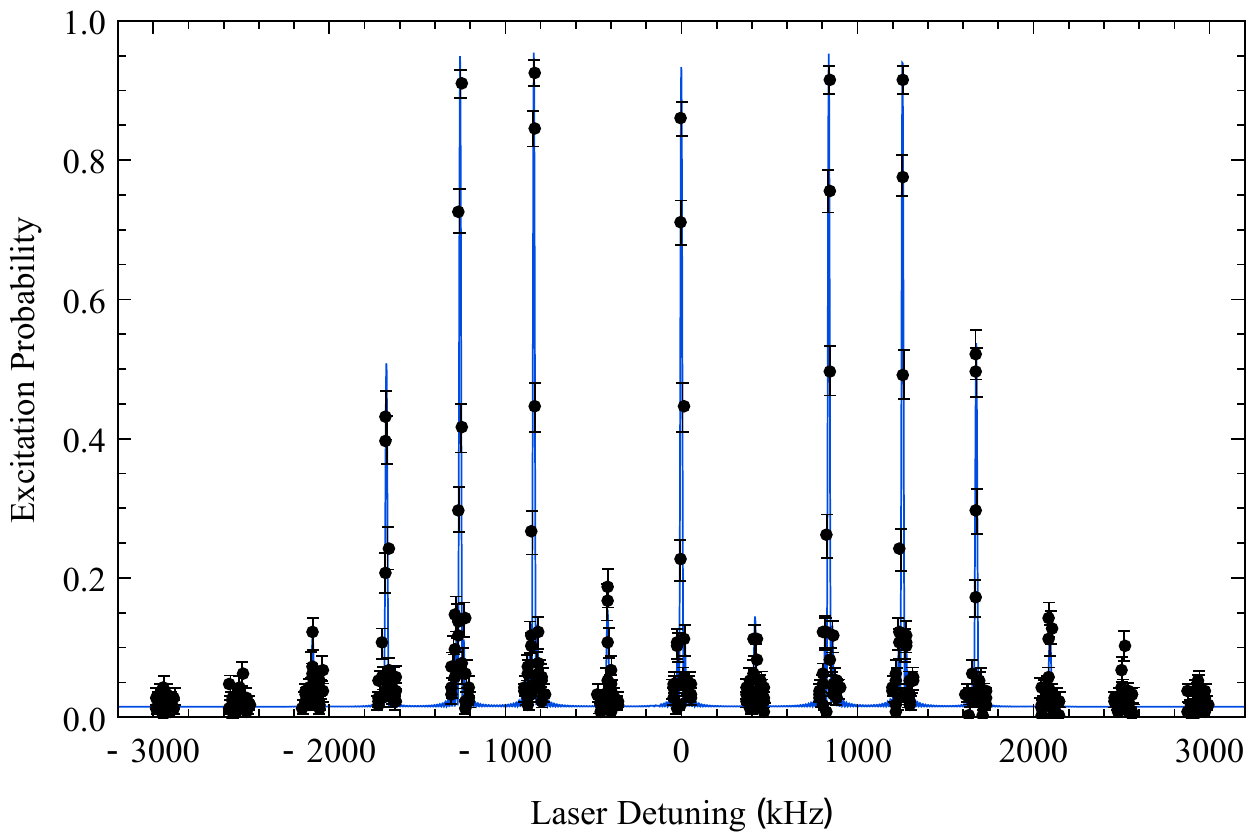}
\caption{}
\vspace*{5mm}
\end{subfigure}
\begin{subfigure}{0.5\textwidth}
\centering
\vspace{-0.35in}\includegraphics[scale=0.55]{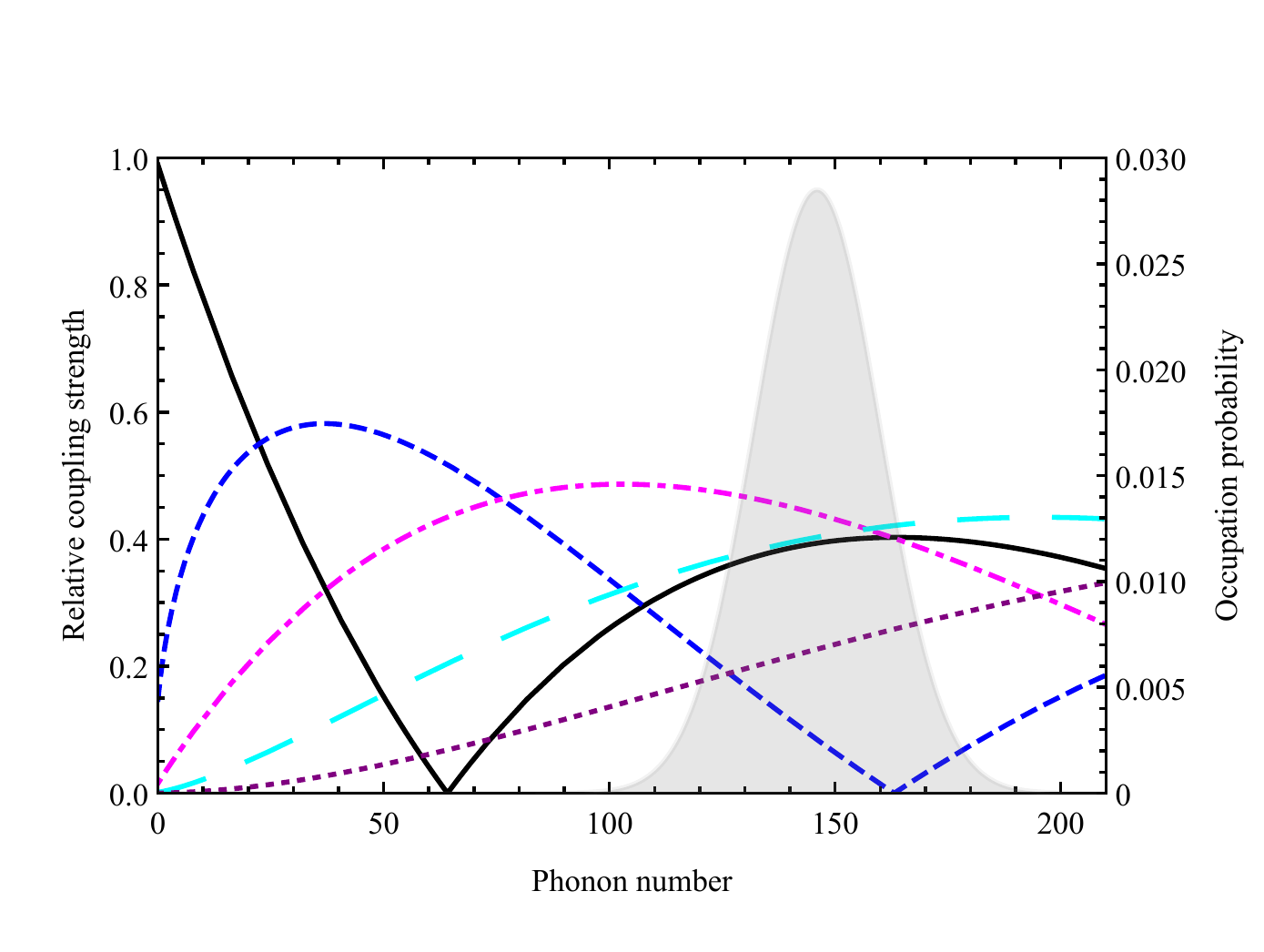}
\vspace{-0.1in}\caption{}
\vspace*{5mm}
\end{subfigure}
\begin{subfigure}{0.5\textwidth}
\centering
\includegraphics[scale=0.55]{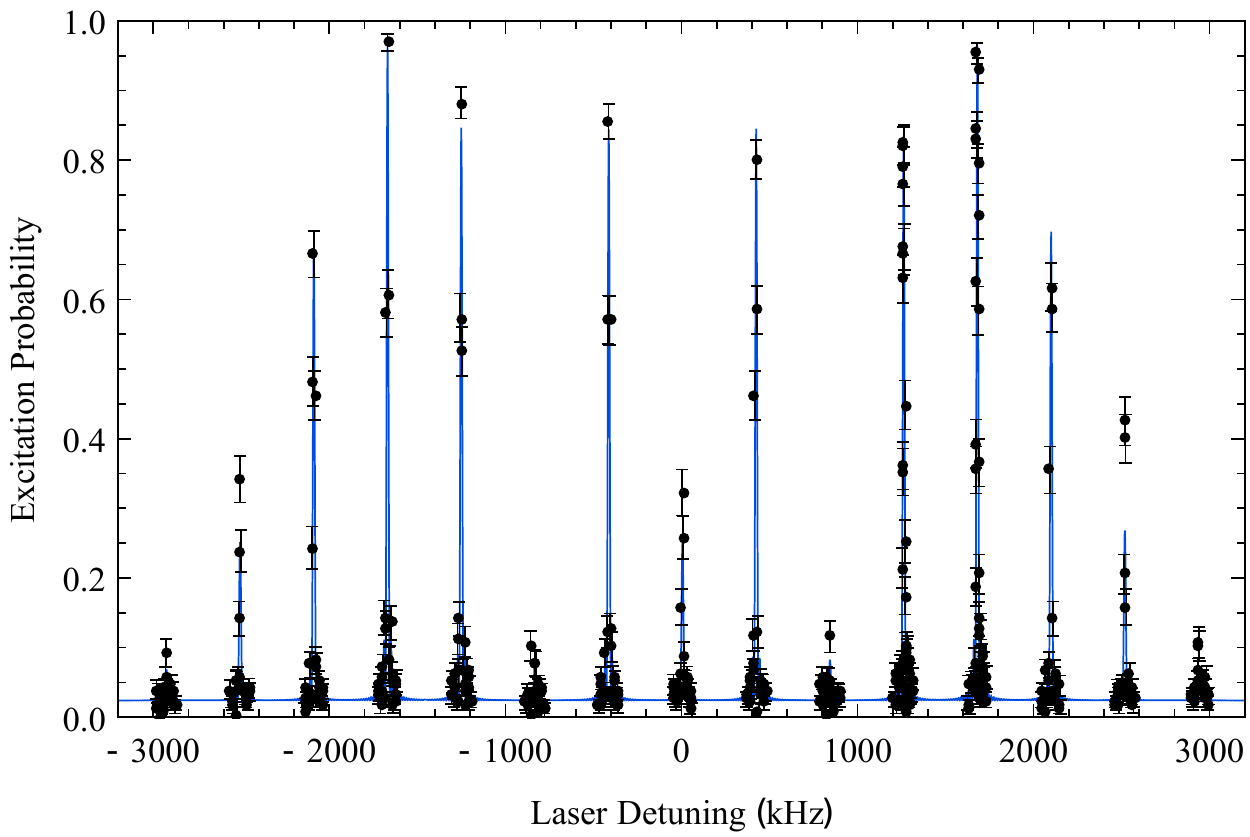}
\caption{}
\end{subfigure}
\begin{subfigure}{0.5\textwidth}
\centering
\vspace{-0.35in}\includegraphics[scale=0.55]{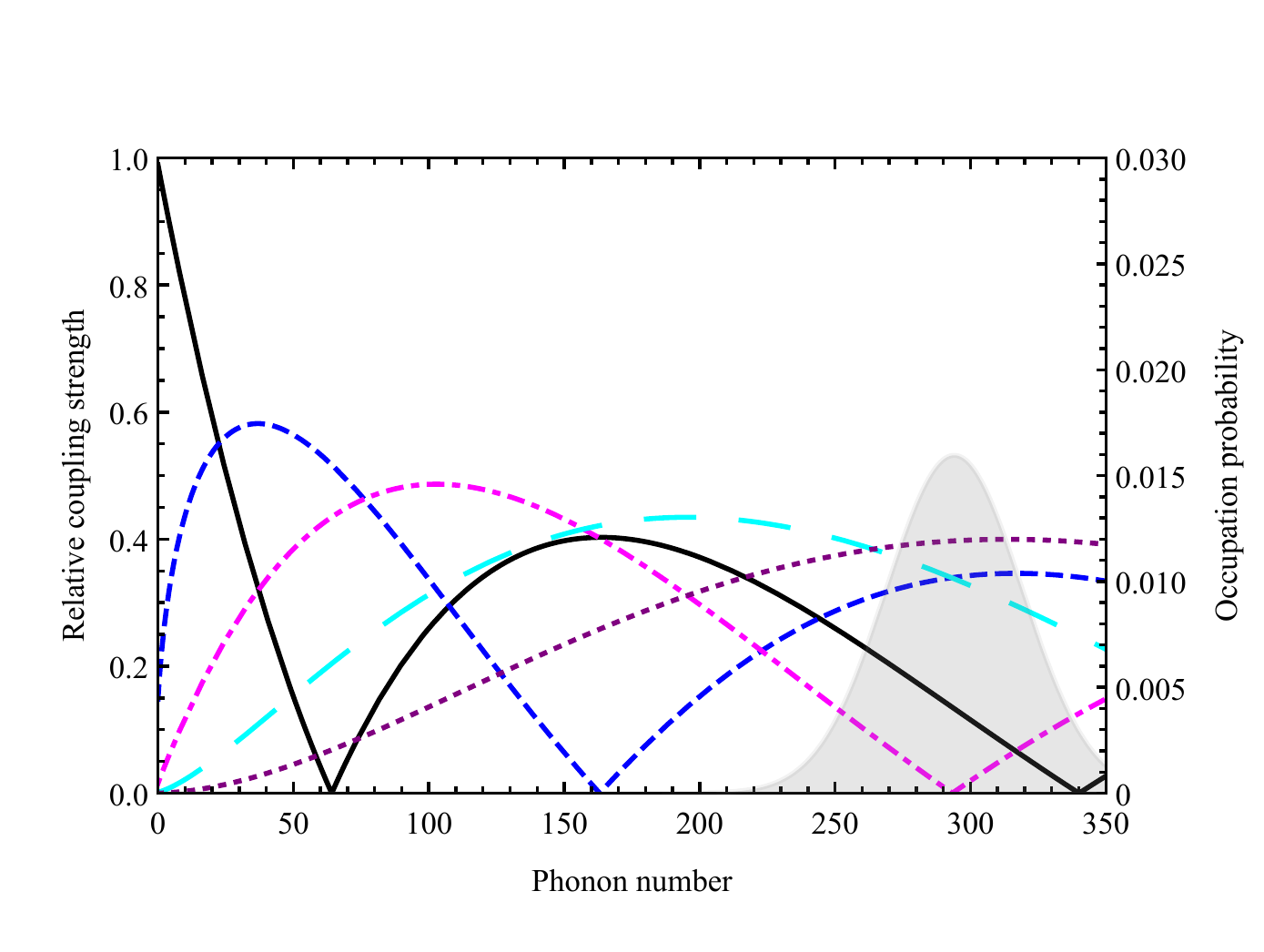}
\vspace{-0.1in}\caption{}
\end{subfigure}
\caption{\textbf{(a)} Spectrum after $\SI{10}{\milli\second}$ of sideband cooling followed by $\SI{10}{\milli\second}$ of sideband heating on the first blue sideband. A fit to the data points finds an average phonon number $\bar{n}=146(2)$ with a standard deviation $\sigma_n=14(1)$ for a distribution assumed to be Gaussian. The Rabi frequency is about $\SI{18}{\kilo\hertz}$. Figure 4\textbf{(a)} is adapted from \cite{joshi19}. \textbf{(b)} Relative coupling strength of the carrier (solid black), first (dashed, blue), second (dot-dashed, magenta), third (long dashes, cyan) and fourth (dotted, purple) blue sidebands as a function of the phonon number. The strengths are calculated according to eq. \ref{rabiCoupEq} for a trap frequency of 420 kHz.  The plot also shows the  probability distribution of the motional state (shaded grey area) as deduced from the sideband spectrum. A coupling strength of 1 corresponds to the Rabi frequency for the free particle (not trapped) which is very close to the coupling strength of the carrier transition in the motional ground state. \textbf{(c)} Spectrum after $\SI{10}{\milli\second}$ of sideband cooling followed by $\SI{10}{\milli\second}$ of sideband heating on the first blue sideband and $\SI{20}{\milli\second}$ on the second blue sideband. A fit to the data points finds an average phonon number $\bar{n}=294(2)$ with a standard deviation $\sigma_n=25(3)$ for a distribution assumed to be Gaussian. The Rabi frequency is about $\SI{16}{\kilo\hertz}$. \textbf{(d)} Similar to \textbf{(b)} but with the distribution corresponding to the sideband spectrum in \textbf{(c)}.}
\label{SBH}
\end{figure}

Starting from a ground state cooled ion, the 729 nm laser is tuned to the frequency of the first blue sideband and the 854 nm `repumper' laser is turned on. The ion is continuously excited to higher motional states until the first zero coupling point for the first sideband is reached. At this point, the motional state of the ion is described by a statistical distribution of Fock states, which can be well approximated by a Gaussian \cite{joshi19}, close to the zero coupling point. The evolution of the ion's motional state during sideband heating can be simulated and we can calculate the probability of the ion to be in a particular state $\ket{n}$ as a function of time.  This is represented in fig. \ref{twoDPlot}, where it can be seen that the motional state's population remains narrowly distributed and converges at particular values where the light-ion coupling vanishes. This particular state of the ion gives rise to a characteristic spectrum (shown in fig \ref{SBH}(a)) where the first order sidebands are heavily suppressed whereas the higher order sidebands remain high, which is well understood by observing the coupling strengths for the different sidebands in fig. \ref{SBH}(b).

\begin{figure}                  
\begin{center}
    \includegraphics[scale=0.5]{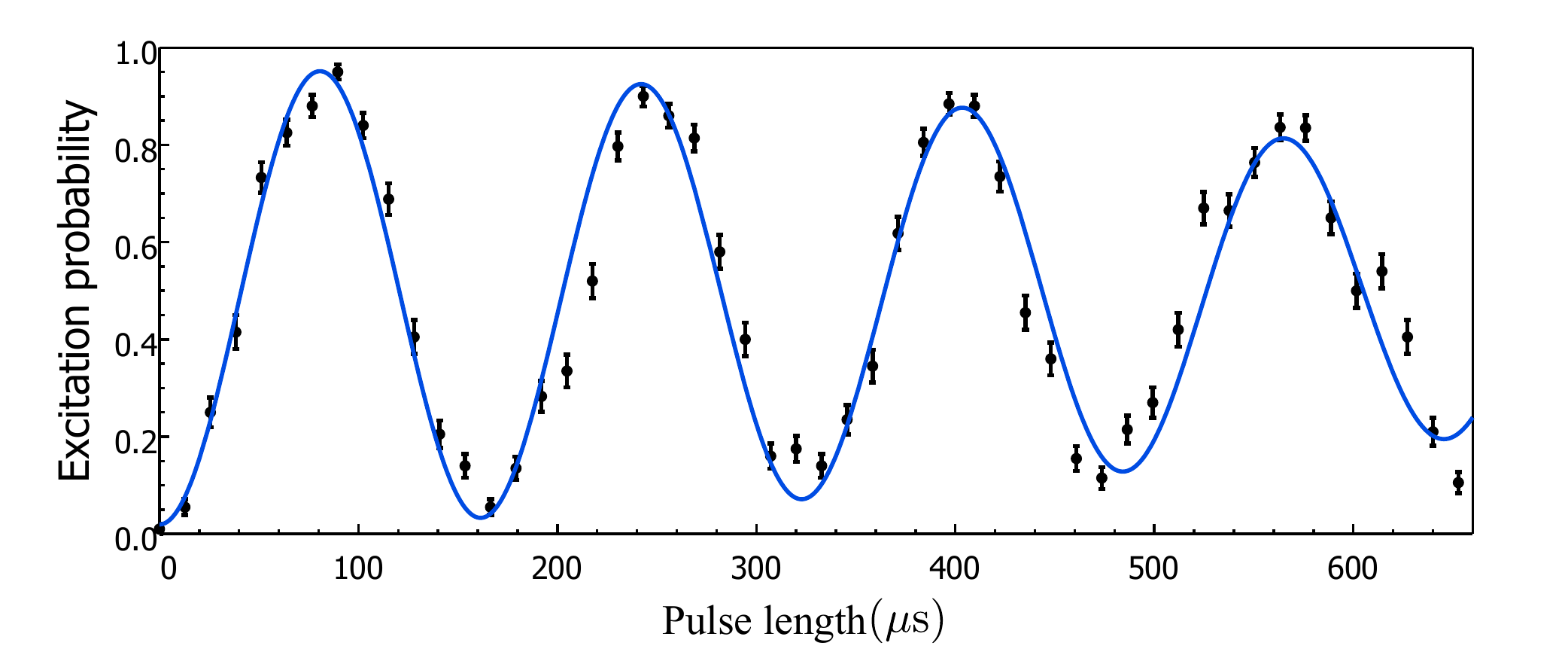}
    \caption{Rabi oscillations on the fourth red sideband after two-stage sideband heating on the first and second blue sidebands. The fit assumes a Gaussian distribution of the motional states with the parameters found from the spectrum in figure \ref{SBH}(c) and a Gaussian decay of the oscillations (due to decoherence). The Rabi frequency is about $\SI{16}{\kilo\hertz}$ and the decay time constant is $\tau_d=\SI{0.73(7)}{\milli\second}$.}
\label{rabi4RSB}
\end{center}
\end{figure}

The ion can be further excited to higher motional states by addressing a higher order sideband, as can be seen in figure \ref{SBH}(c). This spectrum was taken after irradiating with the 729 nm laser for 10 ms tuned to the first blue sideband followed by 20 ms on the second blue sideband. Assuming a Gaussian distribution of the ion's motional state, fitting the data in fig. \ref{SBH}(c) yields $\bar{n}=294(2)$ with a standard deviation $\sigma_n=25(3)$. Although the spread of this distribution is not negligible, we are still able to drive the ion coherently  on certain sidebands and observe Rabi oscillations, as we show in figure \ref{rabi4RSB} where the 729 nm laser is tuned to the fourth red sideband. This is possible thanks to the limited variation in this sideband's coupling strength over the range spanned by the ion's motional state distribution. It is noteworthy that there is almost a full transfer from ground to excited state on the first flop of the Rabi oscillations. The contrast of the oscillations decays for longer pulse times because of the combined effect of the spread of the probability distribution of the motional state and fluctuations of the Rabi frequency stemming from laser intensity noise.

\subsection{Quantum superposition of high motional states}\label{superHighN} 
Even though the ion is left in a incoherent mixture of motional states after sideband heating, we will, for now, consider for simplicity that the ion is in a particular Fock state $\ket{g,n}$. After the heating stage, we wish to prepare the ion in a coherent superposition of motional states in order to measure the coherence time, like in section \ref{GS}. There is however a major difference between the present situation and the case of a ground-state cooled ion, which prevents us from using the same technique. Indeed, the method presented in section \ref{GS} relied on the fact that the population in $\ket{g,0}$ was not affected by a pulse on a red sideband. This is not applicable with high motional states: assume we start with an ion in $\ket{g,n}$ and apply a $\pi/2$ pulse on the carrier followed by a $\pi$ pulse on the third red sideband (for instance). The final state will be $(1/\sqrt 2)( \ket{g,n+3} + \ket{e,n-3} )$. The ion is in a superposition of both motional and electronic (internal) states. This superposition is affected chiefly by the decoherence of the electronic superposition which is much faster that the motional one. 

The interference pattern from a Ramsey experiment using this superposition would lose all contrast after a millisecond or so (our optical coherence time) and therefore cannot be used to measure the motional coherence time. Instead we apply, for instance, a $\pi/2$ pulse on the carrier followed by a $\pi/2$ pulse on the third red sideband. This creates a superposition of $\ket{e,n-3}$, $\ket{g,n}$, $\ket{e,n}$ and $\ket{g,n+3}$. After a wait time $T$ where the lasers are off, a $\pi/2$ pulse on the third red sideband followed by a $\pi/2$ pulse on the carrier completes the sequence. Making the approximation that the Rabi frequency of the carrier is the same for the motional states $\ket{n-3}$, $\ket{n}$ and $\ket{n+3}$, the probability of excitation at the end of the sequence is given by (see Appendix \ref{math})
\begin{align}
    P_e=&\frac{1}{8}\left[4-2V_{\text{mot}}(T)\cos\left(3\omega_z\left(T+2\tau_r\right)\right)+V_{\text{opt}}(T)\left(\cos\left(6\omega_z\left(T+\tau_r\right)\right)+\cos\left(6\omega_z\tau_r\right)\right)\right]
    \label{highNProba}
\end{align}
where $\tau_r$ is the duration of the sideband pulse and we have used the fact that the frequency difference between the carrier and the sideband is $3\omega_z$.  Here again, we have added visibility factors $V_{\text{mot}}(T)$ and $V_{\text{opt}}(T)$ which do not appear in the derivation, to take into account decoherence. This equation contains two time-dependent terms: one oscillating with a frequency $3\omega_z$ and one with a frequency $6\omega_z$, which correspond to the interferences of the populations in states separated by 3 and 6 quanta of motion (phonons) respectively. That is, for the term in $3\omega_z$, the pairs are $\ket{g,n}$, $\ket{g,n+3}$ and $\ket{e,n-3}$, $\ket{e,n}$. For the term in $6\omega_z$, the interfering pair is $\ket{e,n-3}$ and $\ket{g,n+3}$. These two states correspond to different internal electronic states of the ion and therefore this superposition is subject to optical decoherence. We therefore expect the visibility $V_{\text{opt}}(T)$ of the oscillations at $6\omega_z$ to vanish after a time corresponding to the optical coherence time of our system. This is not the case for the term oscillating at $3\omega_z$ since the pairs of interfering populations are in the same electronic state.

\begin{figure}               
\begin{subfigure}{0.5\textwidth}
\centering
\includegraphics[scale=0.55]{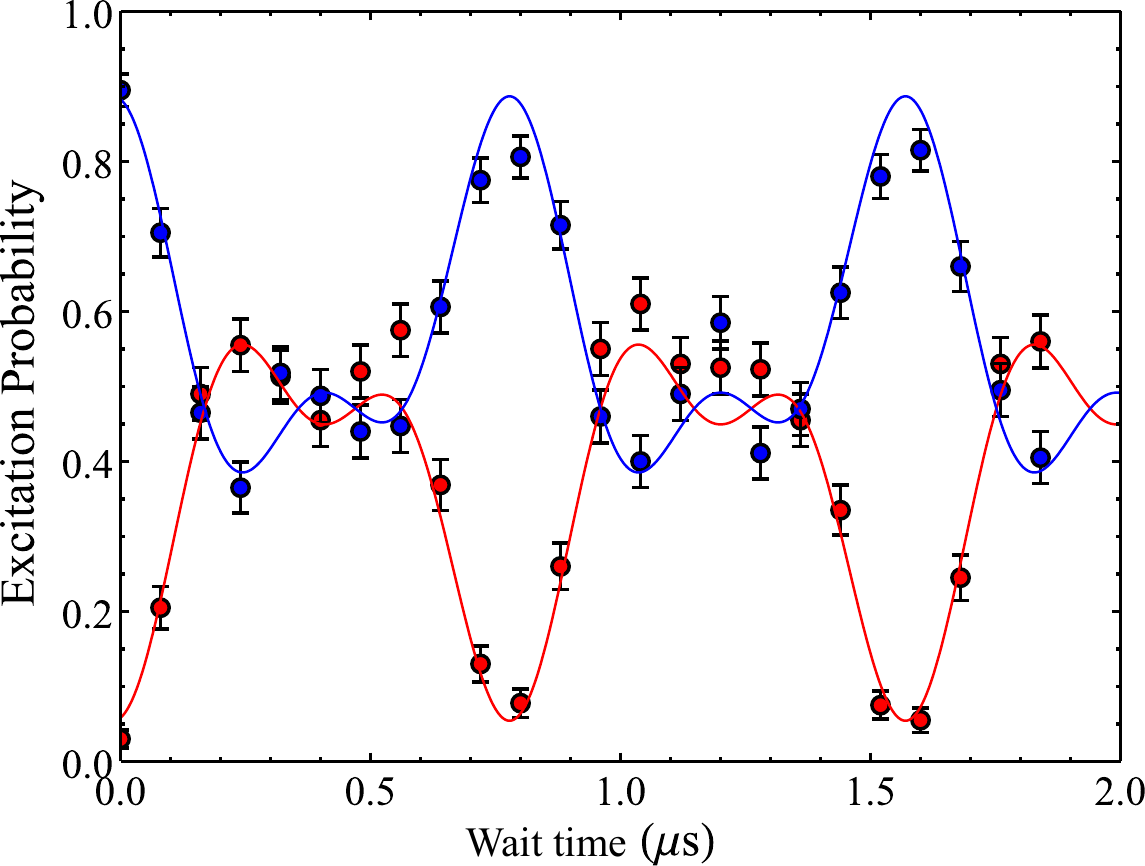}
\caption{}
\vspace{0.5em}
\end{subfigure}
\begin{subfigure}{0.5\textwidth}
\centering
\includegraphics[scale=0.55]{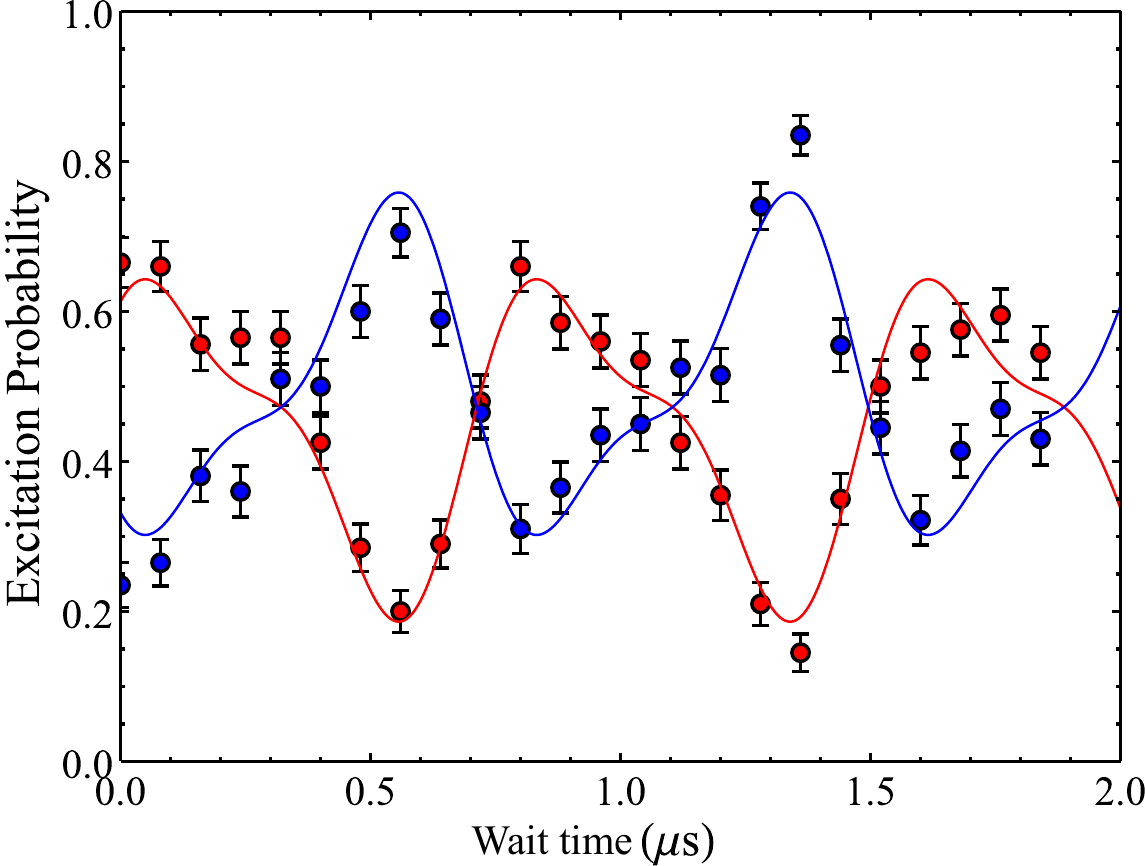}
\caption{}
\vspace{0.5em}
\end{subfigure}
\begin{subfigure}{0.5\textwidth}
\centering
\includegraphics[scale=0.55]{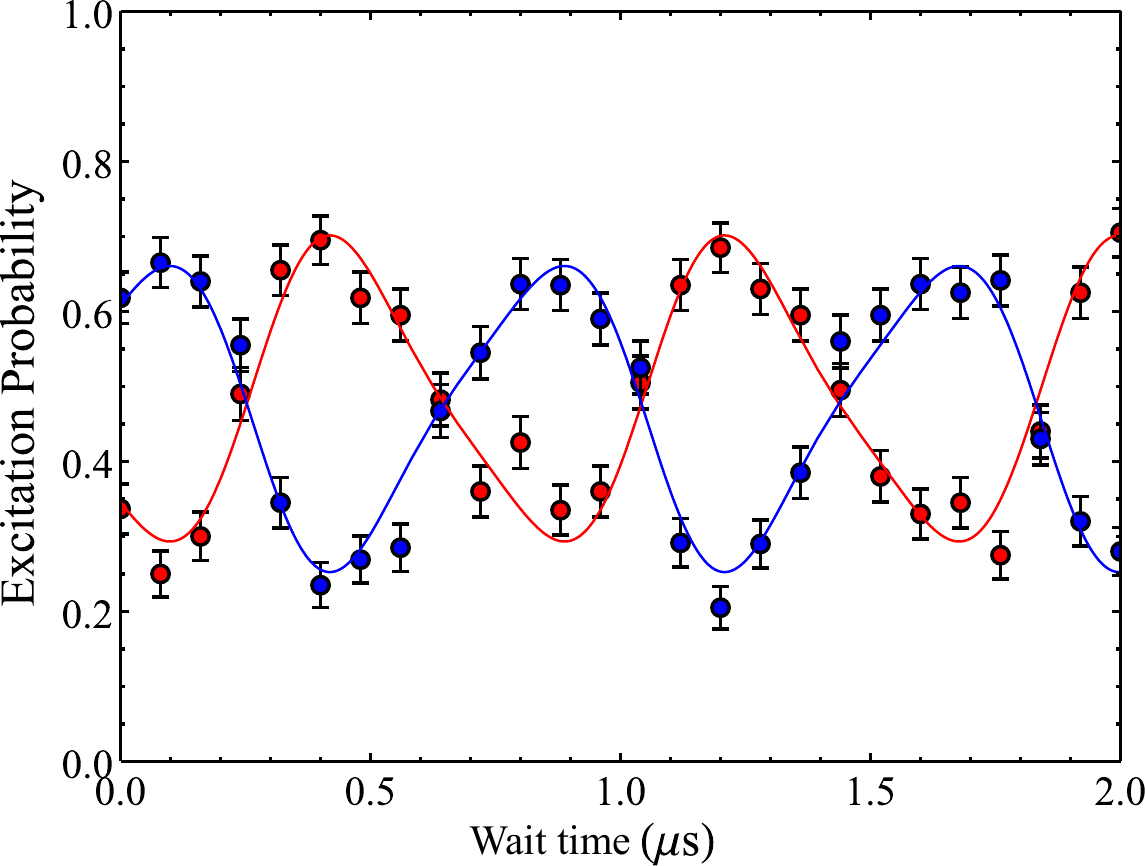}
\caption{}
\end{subfigure}
\begin{subfigure}{0.5\textwidth}
\centering
\includegraphics[scale=0.55]{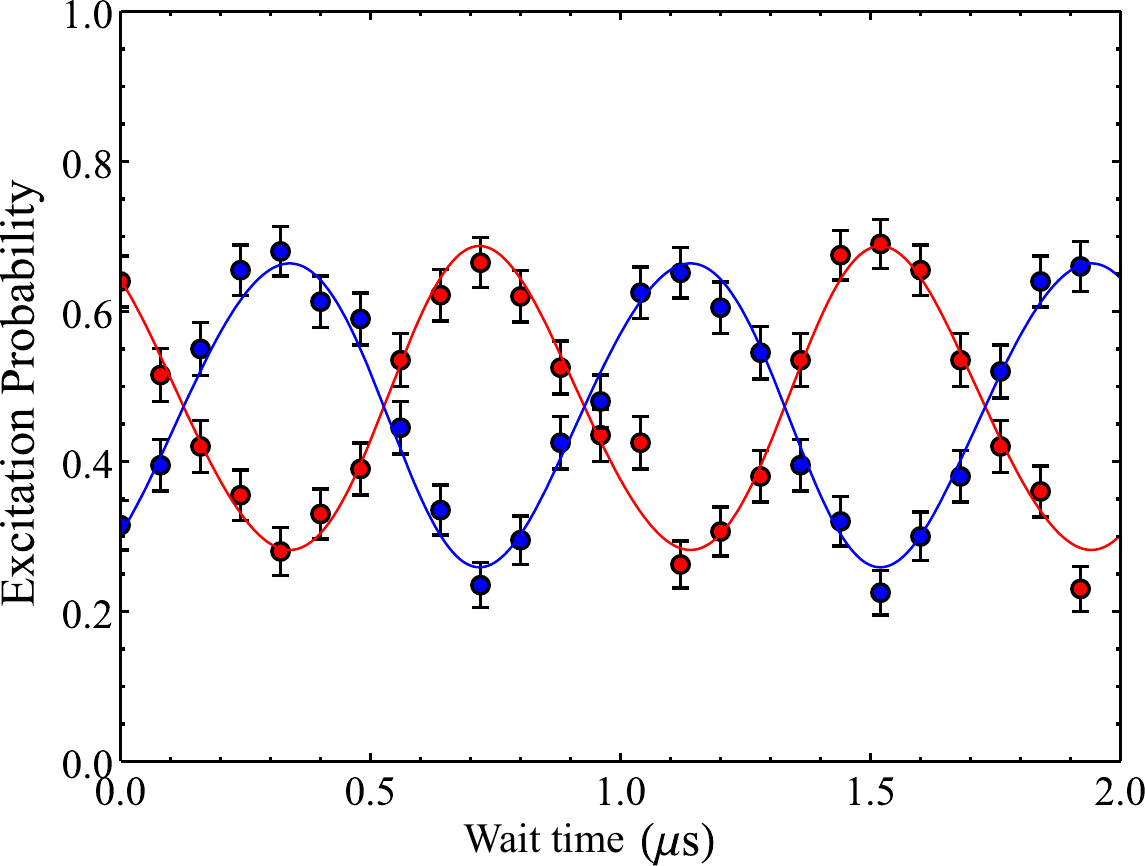}
\caption{}
\end{subfigure}
\caption{Ramsey experiments with a superposition of $\ket{e,n-3}$, $\ket{g,n}$, $\ket{e,n}$ and $\ket{g,n+3}$ with initial wait times $T_0$ of \textbf{(a)} $\SI{10}{\micro\second}$, \textbf{(b)} $\SI{600}{\micro\second}$, \textbf{(c)} $\SI{1}{\milli\second}$ and \textbf{(d)} $\SI{20}{\milli\second}$. The experiment contains two interleaved sequences differing by a $\pi$ offset on the final laser pulse resulting in the two curves shown here.}
\label{HighNFreePrec}
\end{figure}

Figure \ref{HighNFreePrec} gives examples of the results obtained by performing the sequence described above for different wait times $T$. An interference pattern described by equation \ref{highNProba} is obtained by scanning the wait time around a certain value $T_0$: $T=T_0+\delta T$. For a wait time much shorter than the optical coherence time, the two frequency components are clearly visible (Fig. \ref{HighNFreePrec}(a)) and the visibility is maximal. The high frequency component of the oscillation fades away as the wait time $T$ approaches the optical coherence time (Fig. \ref{HighNFreePrec}(b) and (c)) and completely vanishes for a much longer wait time (Fig. \ref{HighNFreePrec}(d)). At this point, one frequency component remains and the interference pattern is described by a simple sinusoid with a maximum visibility of one half.

\begin{figure}               
\begin{center}
\includegraphics[scale=0.65]{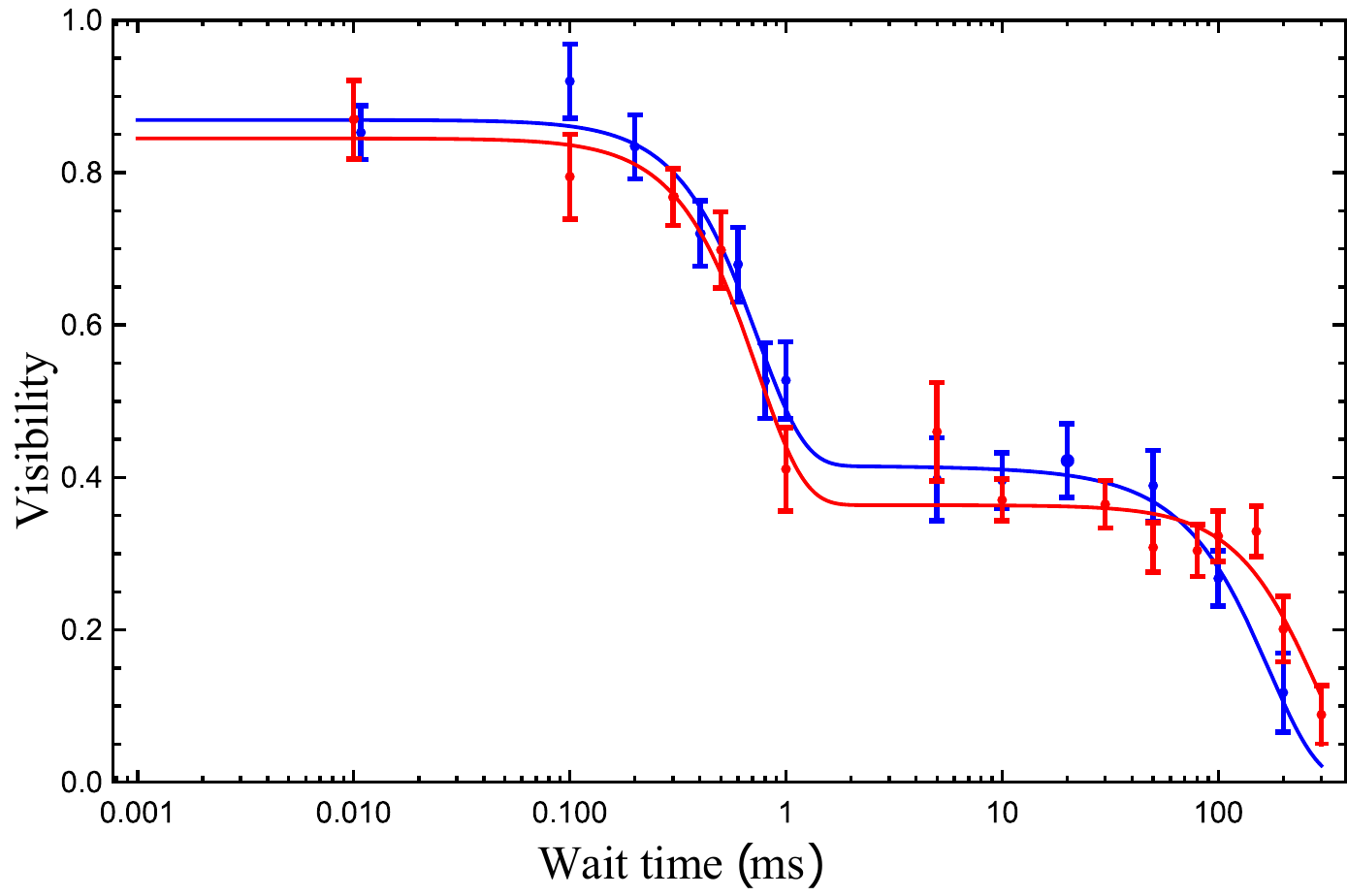}
\caption{Visibility in Ramsey experiments as a function of the wait time for a superposition of $\ket{e,n-3}$, $\ket{g,n}$, $\ket{e,n}$ and $\ket{g,n+3}$ (blue curve) and $\ket{e,n-3}$, $\ket{e,n-1}$,  $\ket{g,n}$ and $\ket{g,n+2}$ (red curve).}
\label{HighNVis03}
\end{center}
\end{figure}

We define the overall visibility of the oscillations $V(T)$ as the average of $V_{\text{mot}}(T)$ and $V_{\text{opt}}(T)$; it is plotted as a function of the wait time $T$ in figure \ref{HighNVis03} for a superposition of $\ket{e,n-3}$, $\ket{g,n}$, $\ket{e,n}$ and $\ket{g,n+3}$. We also show on the same figure the results obtained for a superposition of $\ket{e,n-3}$, $\ket{e,n-1}$, $\ket{g,n}$ and $\ket{g,n+2}$. This superposition yields oscillations at $2\omega_z$ and $4\omega_z$ and is obtained by replacing the pulses on the carrier by pulses on the first red sideband. From these plots, the decay of the optical coherence can clearly be seen occurring before the motional decoherence becomes significant (note that the $x$ axis has a  logarithmic scale). The data points can be well fitted using an expression for the visibility made of a sum of two Gaussian functions corresponding to the optical and motional coherences. We do not take account of motional heating explicitly because we have carried out numerical simulations that show a negligible effect of heating on the coherence properties of the statistical mixture of states produced here. This is in contrast to section \ref{GS}, where the effect of heating out of the ground state is much more significant. We therefore write the visibility $V(T)$ in the following way:
\begin{align}
    V(T)=\frac{1}{2}\left(V_{\text{opt}}^0\exp{\left(-\frac{T^2}{\tau^2}\right)}+V_{\text{mot}}^0\exp{\left(-\frac{1}{2}(\Delta n)^2\sigma^2T^2\right)}\right).
\end{align}
The first term in this equation corresponds to the optical coherence which decays with a characteristic time $\tau=\SI{0.75(6)}{\milli\second}$ and $V_{\text{opt}}^0=0.46(2)$ for the $\ket{e,n-3}$, $\ket{g,n}$, $\ket{e,n}$, $\ket{g,n+3}$ superposition and $\tau=\SI{0.74(11)}{\milli\second}$ and $V_{\text{opt}}^0=0.48(4)$ for $\ket{e,n-3}$, $\ket{e,n-1}$,  $\ket{g,n}$ and $\ket{g,n+2}$. For the second term of the equation, corresponding to the motional coherence part, we find standard deviations $\sigma=\SI{2.8(3)}{\radian\per\second}$ and $\sigma=\SI{2.6(3)}{\radian\per\second}$ with initial visibilities $V_{\text{mot}}^0=0.41(2)$ and $V_{\text{mot}}^0=0.36(2)$ for the two superpositions, respectively. These two values are very close to each other and also in good agreement with those found for a ground state cooled ion. The deviations may be explained in part by the fact that the data for the ground state and high motional state were acquired on different days and it is possible that the noise environment had slightly changed. These results indicate nevertheless that the motional coherence time of a superposition does not depend directly on the quantum number of the motional states involved ($n$) but only on their difference ($\Delta n$). 
 
\pagebreak

\section{Conclusion}    \label{conclusion}
We have shown that we are able to create coherent superpositions of motional states of a trapped ion, initially cooled to its ground state, and that the coherence time of this superposition (up to a value of order one second) can be measured by performing a Ramsey experiment. The loss of coherence with time gives us indications about the noise environment of our trap system. A hybrid model was used to measure dephasing, attributed to slow electrical noise, and the ion's heating rate. Furthermore, using a laser pulse sequence akin to a spin echo for the ion's motion, we were able to suppress the effect of slow noise on a motional superposition, thus increasing the coherence time and giving us a measurement of the ion's heating rate, unbiased (or less significantly so) by dephasing. In addition, we used a technique similar to resolved sideband cooling to create highly excited states. From this incoherent state preparation, it is still possible to coherently drive the ion and perform Ramsey-like experiments with the ion in a coherent superposition of motional states. Measurements of the motion's dephasing allowed us to verify that the coherence time depends on the separation of the motional states in the superposition. Indeed, we found that dephasing is largely independent of the excitation state of the ion's motion. Additional data taken for different excitation states would help verify the trend. The experiments presented here would also benefit from a better state preparation with a narrower probability distribution of the motional state. This should be achievable with a carefully tuned sideband heating sequence.

The ability to perform coherent manipulations on ions in highly excited motional states allows motional superpositions to be created with various state spacings, thus allowing us to tune the sensitivity of the superposition to external perturbations. Used in combination with heating rate measurements by sideband thermometry and measurements of technical electrical noise \cite{sedlacek18}, the methods developed here could help to better characterise and understand the sources of motional decoherence (heating and dephasing) in ion traps.  This has long been an issue for quantum information processing \cite{hite13,bruzewicz15}, particularly in micro-fabricated ion traps \cite{talukdar16} where the ions are close to the surface of the trap's electrodes. Systems with low heating rates and long motional coherence times are less susceptible to errors, thus enabling precision measurements, computation and simulation with higher fidelities.   

\section*{Acknowledgements}
The research leading to these results has received funding  from  the  People  Programme  (Marie  Curie  Actions) of the European Union’s Seventh Framework Programme  (FP7/2007-2013)  under  REA  grant  agreement no.  317232. This  work  was  supported  by  the  UK  Engineering and Physical Sciences Research Council (Grants EP/L016524/1 and EP/P024890/1).

\bibliographystyle{unsrt}
\bibliography{bibliography.bib}
\pagebreak

\appendix
\section{Derivations of the excitation probabilities}\label{math}

Here we derive the expression for the probability for the ion to be in its electronic excited state $\ket{e}$ after a motional Ramsey pulse sequence as described in section \ref{GS}. The ion is assumed to be initially in the ground state $\ket{g,0}$. We take the example of a coherent superposition of $\ket{g,0}$ and $\ket{g,1}$ but the reasoning can be generalised to any superposition of $\ket{g,0}$ and $\ket{g,n}$.

The propagator describing the coupling between the states $\ket{e,n+s}$ and $\ket{g,n}$ in the basis $\lbrace\ket{e,n+s},\ket{g,n}\rbrace$ is \cite{leibfried03}:
\begin{align}
    T_n^s(t)=
    \begin{pmatrix}
    \cos\big(\Omega_{n,n+s}t/2\big)&-ie^{i(\phi+\vert s\vert\pi/2)}\sin\big(\Omega_{n,n+s}t/2\big)\\
    -ie^{-i(\phi+\vert s\vert\pi/2)}\sin\big(\Omega_{n,n+s}t/2\big)&\cos\big(\Omega_{n,n+s}t/2\big)
    \end{pmatrix}.
\end{align}
The ion is initially in the state $\ket{\psi}=\ket{g,0}$. We denote the phase of the $j^{th}$ pulse by $\phi_j$. The first pulse is a $\pi/2$ pulse on the carrier of duration $\tau_{\pi/2}=\pi/(2\Omega_{0,0})$. The propagator for this transition is
\begin{align}
    T_0^0=\frac{1}{\sqrt{2}}
    \begin{pmatrix}
    1&-i\\
    -i&1
    \end{pmatrix}
\end{align}
where the phase of the first pulse $\phi_1$ is set to zero. After the first pulse, the state of the ion is
\begin{align}
    \ket{\psi}=\frac{1}{\sqrt{2}}\left(\ket{g,0}-i\ket{e,0}\right).
\end{align}
The second pulse is a $\pi$ pulse on the first red sideband of duration $\tau_{\pi}=\pi/{\Omega_{0,1}}$. The propagator for this transition is
\begin{align}
    T_1^{-1}=
    \begin{pmatrix}
    0&e^{i\phi_2}\\
    -e^{-i\phi_2}&0
    \end{pmatrix}
\end{align}
and after this pulse the state of the ion is
\begin{align}
    \ket{\psi}=\frac{1}{\sqrt{2}}\left(\ket{g,0}+ie^{-i\phi_2}\ket{g,1}\right).
\end{align}
The third pulse is identical to the second one but with phase $\phi_3$,  after which
\begin{align}
    \ket{\psi}=\frac{1}{\sqrt{2}}\left(\ket{g,0}+ie^{i(\phi_3-\phi_2)}\ket{e,0}\right).
\end{align}
The last pulse is, like the first one, a $\pi/2$ pulse on the carrier but with a non-zero phase. The final state of the ion after the four pulses is
\begin{align}
    \ket{\psi}=\frac{1}{2}\left(\left(1+e^{i(\phi_3-\phi_2-\phi_4)}\right)\ket{g,0}+i\left(e^{i(\phi_3-\phi_2)}-e^{i\phi_4} \right)\ket{e,0}\right)
\end{align}
and the probability of excitation is
\begin{align}
\begin{split}
    P_e&=\vert\braket{e,0\vert\psi}\vert^2\\
    P_e&=\frac{1}{2}\left(1-\cos\left(\phi_3-\phi_2-\phi_4\right)\right).
\end{split}
\end{align}
The phases $\phi_{2}$, $\phi_{3}$ and $\phi_{4}$ are phase offsets due to the frequency switching of the direct digital synthesizer used to modulate the laser. We can express them explicitly:
\begin{align}
    \begin{split}
        \phi_2&=\tau_{\pi/2}\left(\omega_c-\omega_r\right)\\
        \phi_3&=\left(\tau_{\pi/2}+T\right)\left(\omega_c-\omega_r\right)\\
        \phi_4&=-2\tau_{\pi}\left(\omega_c-\omega_r\right)
    \end{split}
\end{align}
and rewrite the probability of excitation as:
\begin{align}
    P_e=\frac{1}{2}\left[1-\cos\left(\left(\omega_c-\omega_r\right)\left(2\tau_{\pi}+T\right)\right) \right]
\end{align}
and in this case, $\omega_c-\omega_r=\omega_z$, so
\begin{align}
    P_e=\frac{1}{2}\left[1-\cos\left(\omega_z\left(2\tau_{\pi}+T\right)\right)\right]
\end{align}
which corresponds to equation \ref{proba0n}.

Similarly, we can derive the probability of excitation after the pulse sequence described in section \ref{highN}. We assume the ion is in the state $\ket{\psi}=\ket{g,n}$. The first pulse is a $\pi/2$ pulse on the carrier of duration $\tau_c$. Like the previous case, the first pulse creates the superposition
\begin{align}
    \ket{\psi}=\frac{1}{\sqrt{2}}\left(\ket{g,n}-i\ket{e,n}\right).
\end{align}
The second pulse is a $\pi/2$ pulse on the third red sideband of duration $\tau_r$. The propagator in both the bases $\lbrace\ket{e,n},\ket{g,n+3}\rbrace$ and $\lbrace\ket{e,n-3},\ket{g,n}\rbrace$ is
\begin{align}
    T_n^{-3}=\frac{1}{\sqrt{2}}
    \begin{pmatrix}
    1&-e^{i\phi_2}\\
    e^{-i\phi_2}&1
    \end{pmatrix}
\end{align}
and the state of the ion after the second pulse is
\begin{align}
    \ket{\psi}=\frac{1}{2}\left(-e^{i\phi_2}\ket{e,n-3}+\ket{g,n}-i\ket{e,n}-ie^{-i\phi_2}\ket{g,n+3}\right).
\end{align}
The third pulse is identical to the second one and puts the ion in the state
\begin{align}
\begin{split}
    \ket{\psi}=&\frac{1}{2\sqrt{2}}\bigg[-\left(e^{i\phi_2}+e^{i\phi_3}\right)\ket{e,n-3}+\left(1-e^{i(\phi_2-\phi_3)}\right)\ket{g,n}\\
    &-i\left(1-e^{-i(\phi_2-\phi_3)}\right)\ket{e,n}-i\left(e^{-i\phi_2}+e^{-i\phi_3}\right)\ket{g,n+3}\bigg].
\end{split}
\end{align}
The last pulse is a pulse on the carrier transition. The duration of the pulse is calibrated to be a $\pi/2$ for the transition $\ket{g,n}\leftrightarrow\ket{e,n}$ but we will assume that the pulse is also a $\pi/2$ pulse for the transitions $\ket{g,n-3}\leftrightarrow\ket{e,n-3}$ and $\ket{g,n+3}\leftrightarrow\ket{e,n+3}$. In other words, we assume that the Rabi frequency of the carrier transition does not change on the interval $[n-3,n+3]$. The state of the ion after the last pulse is
\begin{align}
    \begin{split}
        \ket{\psi}=&\frac{1}{4}\bigg[-ie^{-i\phi_4}\left(e^{i\phi_2}+e^{i\phi_3}\right)\ket{g,n-3}-\left(e^{i\phi_2}+e^{i\phi_3}\right)\ket{e,n-3}\\
        &+\left(1-e^{i\left(\phi_2-\phi_3\right)}-e^{-i\phi_4}\left(1-e^{-i(\phi_2-\phi_3)}\right)\right)\ket{g,n}\\
        &-i\left(1-e^{-i\left(\phi_2-\phi_3\right)}+e^{i\phi_4}\left(1-e^{i(\phi_2-\phi_3)}\right)\right)\ket{e,n}\\
        &-i\left(e^{-i\phi_2}+e^{-i\phi_3}\right)\ket{g,n+3}-e^{i\phi_4}\left(e^{-i\phi_2}+e^{-i\phi_3}\right)\ket{e,n+3}\bigg].
    \end{split}
\end{align}
The probability of excitation will be the sum of the probabilities of the ion being in $\ket{e,n-3}$, $\ket{e,n}$, $\ket{e,n+3}$. We get:
\begin{align}
    \begin{split}
        &P_e(n-3)=\frac{1}{8}\left(1+\cos\left(\phi_2-\phi_3\right)\right)\\
        &P_e(n)=\frac{1}{8}\left(2+\cos\left(\phi_4\right)-2\cos(\phi_2-\phi_3)-2\cos\left(\phi_4+\phi_2-\phi_3\right)+\cos\left(\phi_4+2\phi_2-2\phi_3\right)\right)\\
        &P_e(n+3)=\frac{1}{8}\left(1+\cos\left(\phi_2-\phi_3\right)\right).
    \end{split}
\end{align}
Overall, the probability of excitation is:
\begin{align}
    \begin{split}
        P_e=&P_e(n-3)+P_e(n)+P_e(n+3)\\
        P_e=&\frac{1}{8}\left[4+\cos\left(\phi_4\right)-2\cos\left(\phi_4+\phi_2-\phi_3\right)+\cos\left(\phi_4+2\phi_2-2\phi_3\right)\right].
    \end{split}
\end{align}
As for the low $n$ case, we replace the phases by their explicit expressions:
\begin{align}
    \begin{split}
        &\phi_2=\left(\omega_c-\omega_r\right)\tau_c\\
        &\phi_3=\left(\omega_c-\omega_r\right)\left(\tau_c+T\right)\\
        &\phi_4=-2\left(\omega_c-\omega_r\right)\tau_r
    \end{split}
\end{align}
and using $\omega_c-\omega_r=3\omega_z$, we obtain
\begin{align}
    P_e=&\frac{1}{8}\left[4-2\cos\left(3\omega_z\left(T+2\tau_r\right)\right)+\cos\left(6\omega_z\left(T+\tau_r\right)\right)+\cos\left(6\omega_z\tau_r\right)\right].
\end{align}
\end{document}